\documentstyle[multicol,prb,aps]{revtex}

\def\bra{\langle}
\def\ket{\rangle}

\def\e{\,{\rm e}}

\def\dfrac{\displaystyle\frac}

\def\vk{{\bf k}}
\def\vr{{\bf r}}
\def\vv{{\bf v}}
\def\vx{{\bf x}}

\def\vF{{\bf F}}

\def\psa{{\psi_\alpha}}
\def\psb{{\psi_\beta}}
\def\Psa{{\Psi_\alpha}}
\def\Psb{{\Psi_\beta}}
\def\vol{{\cal V}}
\def\vF{{\rm\bf F}}
\def\vk{{\rm\bf k}}
\def\svk{{\rm k}}
\def\vv{{\rm\bf v}}
\def\vx{{\rm\bf x}}
\def\bl{\big|}
\def\dl{{\delta_l}}
\def\dlp{{\delta_{l+1}}}

\def\dem{{\delta_m}}
\def\dmp{{\delta_{m+1}}}

\def\kp{{k^\prime}}
\def\lp{{l^\prime}}
\def\mp{{m^\prime}}

\def\thph{\theta,\phi}
\def\eab{{E_\alpha-E_\beta}}

\def\solidangle{\int d\Omega_{k\alpha}d\Omega_{k\beta}}
\def\polarangle{\int d\phi_{k\alpha}d\phi_{k\beta}}
\newcommand{\mepv}[1]{\Big\bra\psb\Big|{\frac{\partial V}{\partial #1}}\Big|\psa\Big\ket}
\newcommand{\scap}[1]{\int d\phi(1-\cos\phi)\sigma #1\phi)}
\newcommand{\scat}[1]{\int d\Omega(1-\cos\theta)\sigma #1\theta)}

\begin{document}

\draft

\title{Longitudinal Force on a Moving Potential}
\author{Jian-Ming Tang and D. J. Thouless}
\address{Department of Physics, University of Washington \\
Box 351560, Seattle WA 98195-1560}
\date{submitted to PRB on 2 July 1998}
\maketitle

\begin{abstract}
We show a formal result of the longitudinal force acting on a moving potential.
The potential can be velocity-dependent, which appears in various interesting
physical systems, such as electrons in the presence of a magnetic flux-line,
or phonons scattering off a moving vortex.
By using the phase-shift analysis, we are able to show the equivalence between
the adiabatic perturbation theory and the kinetic theory for the longitudinal
force in the dilute gas limit.
\end{abstract}


\begin{multicols}{2}

\section{Introduction}

In recent work on the theory of the transverse force on vortex moving
relative to a superfluid\cite{Thouless96,Geller98,Wexler98} we have 
used a technique in which the force is calculated by applying a
pinning potential which is made to move, and the reaction on the
pinning potential is calculated perturbatively.  A number of people
have objected privately that this method does not give the
longitudinal force correctly.  We do not agree with this objection,
since we think that it can be used to determine the longitudinal
force.  For simple problems our method gives results that are in
agreement with those obtained by more familiar methods.  This paper
examines this question for some special cases.  We conclude that any
differences from other perturbative methods which may occur are due to
uncertainties in limiting procedures.

For the component of force transverse to the vortex velocity on a
vortex in a superfluid Thouless, Ao and Niu \cite{Thouless96} (which
we refer to as TAN) found that its value is independent of the pinning
potential with which the vortex is made to move.  The magnitude of
this transverse force per unit length is equal to the circulation of
momentum density at large distances multiplied by the velocity of the
vortex relative to the superfluid, to lowest order in the vortex
velocity.  In ideal superconductors a similar method gave a transverse
force equal to the product of the vortex velocity and the circulation
of canonical momentum density \cite{Geller98}.  However, this result
is controversial \cite{Volovik96,Sonin97,Hallhook98}.  The discrepancy
may come from using different calculation schemes, or it may come from
inadequate handling of boundary conditions in some or all of the
calculations \cite{Wexler98}.  TAN's derivation uses the
time-dependent perturbation theory in contrast to the usual approach
using scattering theory and kinetic equations.  It is known that the
kinetic method and the linear-response theory do not have the same
domains of validity\cite{Kubo95}, which motivates us to examine the
applicability of the force formula more carefully.

In Sec.\ II we consider the effect of a moving short-ranged
spherically symmetric potential on a system of noninteracting
particles, and show that the longitudinal force obtained from the
method of TAN is, to lowest order in the velocity, identical to that
which is obtained from kinetic theory.  In Sec.\ III we consider the
long-ranged Aharonov--Bohm type potential, and again find no difference
in the longitudinal force.  The transverse force seems to be more
sensitive to limiting procedures.

\section{The Longitudinal Force}

The Hamiltonian describing a system of noninteracting particles in a
system which is uniform apart from a spherically symmetric scattering
potential whose center is time-dependent has the form
\begin{equation}
H=\sum_{i=1}^N\left\{-\frac{\nabla_i^2}{2m_0}+V[\vr_i-\vr_0(t)]\right\} \;.
\end{equation}
It is convenient to work with a velocity which is switched on
adiabatically, so that $\dot\vr_0=\vv_V\e^{\gamma t}$.  Since we will
only be considering effects linear in $\vv_V$ the frequency dependence
of the response can be obtained by analytical continuation of the
$\gamma$ dependence into the complex plane.  The method used by TAN
involves calculating the expectation value of the force on the
potential to first order in $\vv_V$, using the instantaneous
eigenstates of $H$ as a basis.  This gives
\begin{eqnarray}
\lefteqn{ \vF=-\sum_\alpha f_\alpha\bra\Psa|(\nabla_0H)|\Psa\ket } 
\label{eq:force}\\
&& +\sum_\alpha f_\alpha
\bigg\bra\Psa\bigg|(\nabla_0H)\frac{i{\cal P}_\alpha}{E_\alpha-H+i\gamma}
\vv_V\cdot\nabla_0+{\rm h.c.}\bigg|\Psa\bigg\ket \;, \nonumber
\end{eqnarray}
where $|\Psa\ket$ is an instantaneous eigenvalue of $H$ with
eigenvalue $E_\alpha$, $f_\alpha$ is the equilibrium probability that
the state is occupied, and ${\cal P}_\alpha$ is the projection
operator off the state $\alpha$.  The first term is not proportional
to the vortex velocity $v_V$, and vanishes if the system has spherical
symmetry.  We can simplify Eq. (\ref{eq:force}) further by substituting
$\nabla_0|\Psa\ket$ with $\nabla_0H$, which is $\nabla_0V$, because
$\nabla_0H$ is the commutator of $\nabla_0$ and $H$,
\begin{equation}
\nabla_0|\Psa\ket=\frac{1}{E_\alpha-H}(\nabla_0V)|\Psa\ket \;.
\end{equation}
This gives the resultant longitudinal force as
\begin{eqnarray}
\vv_V\cdot\vF &=&
\sum_{\alpha\neq\beta}\frac{f_\alpha-f_\beta}{\eab}\Big|\bra\Psb|
\vv_V\cdot(\nabla_0V)|\Psa\ket\Big|^2 \nonumber\\
&& \frac{\gamma}{(\eab)^2+\gamma^2} \;. \label{eq:longi}
\end{eqnarray}
In the limit that $\gamma$ goes to zero, which is the case that we
consider in the rest of this paper, the last term gives a
delta function of the energy, so that the first factor on the right
hand side gives $df/dE$.

For noninteracting particles the sum over states can be replaced by a
sum over single-particle wave functions, with the $f_\alpha$ replaced
by fermionic or bosonic occupation probabilities.
For a central potential the energy-conserving matrix elements in Eq.\
(\ref{eq:longi}) can be expressed in terms of phase shifts, using the
relation between the plane wave and spherical wave expansions,
\begin{equation}
\bra\vr|\psi_\svk\ket=\dfrac{4\pi}{\sqrt{\vol}}\sum_{lm}i^lA_l(kr)
Y_{lm}(\hat{r})Y^*_{lm}(\hat{k}) \;,
\end{equation}
where $\vol$ is the system volume, and the radial wave functions at large distances are
\begin{equation}
A_l(kr)=\e^{i\dl}\big[\cos\dl j_l(kr)-\sin\dl n_l(kr)\big] \;.
\end{equation}
The matrix element which needs to be evaluated for the longitudinal
force given by Eq. (\ref{eq:longi}) is
\begin{eqnarray}
\lefteqn{ \mepv{z}=\frac{(4\pi)^2}{\vol}\sum_{lm}\sum_{\lp\mp}i^{l-\lp}
Y_{\lp\mp}(\hat k_\beta)Y^*_{lm}(\hat k_\alpha) }\nonumber\\
&& \int dr\,r^2A^*_\lp(kr)\frac{dV}{dr}A_l(kr)
\int d\Omega\cos\theta Y^*_{\lp\mp}(\hat r)Y_{lm}(\hat r) \;, \label{eq:pot}
\end{eqnarray}
if $\dot\vr_0$ is taken to be in the $\hat z$ direction.
The last angular integral can be carried out by using the following recurrence relation,
\begin{eqnarray}
\cos\theta\,Y_{lm}(\thph)
& = & \sqrt{\frac{(l+m+1)(l-m+1)}{(2l+3)(2l+1)}}Y_{l+1,m}(\thph) \nonumber\\
& + & \sqrt{\frac{(l+m)(l-m)}{(2l+1)(2l-1)}}Y_{l-1,m}(\thph) \;,
\end{eqnarray}
where the orbital angular momenta are changed by one because the gradient is a tensor operator of rank one.
Taking the square of its modulus and integrating $\hat{k}_\alpha$ and $\hat{k}_\beta$ over all possible directions
will get rid of the remaining spherical harmonics, and end up with a summation over orbital angular momenta,
\begin{eqnarray}
\lefteqn{ \solidangle\bigg|\mepv{z}\bigg|^2 }\\
& = & \frac{(4\pi)^4}{\vol^2}\sum_{l}\frac{2}{3}(l+1)\bigg|\int dr\,r^2 A^*_{l+1}\frac{dV}{dr}A_l\bigg|^2 \;. \nonumber
\end{eqnarray}
The summation over magnetic quantum numbers can be easily carried out because there is only one independent direction left after the integration.
The radial integral gives the difference between phase shifts (see Appendix A),
\begin{eqnarray}
\lefteqn{ 2m_0\int drr^2A^*_{l+1}(kr)\frac{dV}{dr}A_l(kr) }\nonumber\\
& = & -\e^{i(\dl-\dlp)}\sin(\dlp-\dl) \;.
\end{eqnarray}
There is no contribution to the force from bound states because their wave functions vanish at large distances, and resonance states may potentially play an important role here.
So the longitudinal force can be written as an integral over all scattering states,
\begin{equation}
\frac{F}{v_V}=N\int dk\frac{\partial f_k}{\partial k}\frac{2k^2}
{3\pi}\sum_l(l+1)\sin^2(\dlp-\dl) \;.
\label{eq:3dlongi}\end{equation}
This result was obtained in a similar way by B\"onig and Sch\"onhammer
\cite{Bonig89}.

There is a very similar calculation for scattering by an axially
symmetric potential, for which the wave functions are
\begin{equation}
\bra\vx|\psi_\svk\ket=\dfrac{1}{\sqrt{\vol}}\sum_{m}i^mA_m(k\rho)
\e^{im(\phi-\phi_{\hat{k}})} \;.
\end{equation}
The calculation for the matrix element is almost the same except for a
different normalization constant,
\begin{eqnarray}
\lefteqn{ \mepv{x} } \nonumber\\
&=& \frac{i\pi}{\vol}\sum_m\e^{-im\phi_\alpha}\int d\rho\rho\frac{dV}
{d\rho}A_m \\
&& \Big(-\e^{i(m+1)\phi_\beta}A^*_{m+1}+\e^{i(m-1)\phi_\beta}A^*_{m-1}\Big) \;. \nonumber
\end{eqnarray}
After integrating all incident directions,
only the radial integral remains,
\begin{eqnarray}
\lefteqn{ \polarangle\bigg|\mepv{x}\bigg|^2 }\nonumber\\
& = & \frac{(2\pi)^3\pi}{\vol^2}\sum_m\bigg|\int d\rho\rho A^*_{m+1}
\frac{dV}{d\rho}A_m\bigg|^2 \;,
\label{eq:forcesquare}\end{eqnarray}
and this gives the connection to the phase shifts in the same fashion,
\begin{eqnarray}
\lefteqn{2m_0\int d\rho\rho A^*_{m+1}\frac{dV}{d\rho}A_m }\nonumber\\
& = & -\frac{2k}{\pi}\e^{i(\dem-\dmp)}\sin(\dmp-\dem) \;.
\end{eqnarray}
This gives the longitudinal force as
\begin{equation}
\frac{F}{Lv_V}=N\int dk\frac{\partial f_k}{\partial k}\frac{k^2}
{2\pi}\sum_m\sin^2(\dmp-\dem) \;, \label{eq:2dlongi}
\end{equation}
where $L$ is the length of the system in the axial direction.

Equations (\ref{eq:3dlongi}) and (\ref{eq:2dlongi}) are identical to
the equations obtained from the standard kinetic theory argument. In
kinetic theory the usual procedure is to keep the scattering potential
fixed and to shift the equilibrium distribution so that particles have
an average momentum at large distances given by an undetermined
multiplier $\vv_V$.  The force is the momentum transfer per unit time
from  particles to the scattering center. In $d$ dimensions this is
related to the differential scattering cross section $\sigma(\vk,\theta)$ by
\begin{eqnarray}
F &=& \hat{\vv}_V\cdot\int\frac{d^dk}{(2\pi)^d}(\vk-m_0\vv_V)
\frac{N}{m_0}f_\svk\bl\vk-m_0\vv_V\bl \\
&& \scat{(\vk-m_0\vv_V,} \nonumber\\
&\simeq& Nv_V\int dkk^{d+1}\frac{\partial f_k}{\partial k} \\
&& \int\frac{d\Omega_\svk}{(2\pi)^d}\cos^2\theta_\svk\scat{(k,} \;, \nonumber
\end{eqnarray}
where, on the right side of the equation, only the first order term in
the velocity has been kept.  The angular integral over all incident
directions can be carried out independently because of the isotropy,
and gives
\begin{equation}
\int d\Omega_\svk\cos^2\theta_\svk=\frac{2\pi^{d/2}}{\Gamma(d/2)d} \;.
\end{equation}
The last scattering cross section integral is often called the
transport  cross section, $\sigma_{tr}$. For a spherically symmetric
potential in three dimensions the scattering amplitude can be expanded
in Legendre polynomials as
\begin{equation}
{\cal F}(\theta)=\frac{1}{k}\sum_l(2l+1)\e^{i\dl}\sin\dl P_l(\cos\theta) \;,
\end{equation}
where $\dl$ is the phase shift for the $l$-th partial wave.
The differential cross section is just the modulus squared of the scattering
amplitude, and the transport cross section is 
\begin{eqnarray}
\sigma_{tr} &=& \scat{(} \nonumber\\
&=& \frac{4\pi}{k^2}\sum_l(l+1)\sin^2(\dlp-\dl) \;. \label{eq:tcs3d}
\end{eqnarray}
This is in agreement with Eq.\ (\ref{eq:3dlongi}).

In two dimensions, the scattering amplitude is expanded in Fourier series,
\begin{equation}
{\cal F}(\phi)=\sqrt{\frac{2i}{\pi k}}\sum_m\e^{i\dem}\sin\dem\e^{im\phi} \;,
\end{equation}
and then transport cross section follows similarly,
\begin{eqnarray}
\sigma_{tr} &=& \scap{(} \nonumber\\
&=& \frac{2}{k}\sum_m\sin^2(\dmp-\dem) \;, \label{eq:tcs2d}
\end{eqnarray}
in agreement with Eq.\ (\ref{eq:2dlongi}).

\section{Long-Ranged Potentials}

For problems such as the scattering of phonons by a quantized vortex
\cite{Pitaevskii58,Iordanskii65} or the scattering of charged particles by an
Aharonov--Bohm flux line \cite{Aharonov59,LL77} we need to consider
potentials which fall off with distance like $1/r^2$ and for which the
phase shifts have a nonzero limit for large angular momentum.  A
potential of the form
\begin{equation}
2m_0V(\rho,k_\phi)=\frac{1}{\rho^2}\bigg(-2i\alpha\frac{\partial}
{\partial\phi}+\beta\bigg) \;. \label{eq:flux}
\end{equation}
This gives an approximate form for scattering of phonons from a quantized
vortex for $\alpha =k\kappa_0/2\pi c$, and for scattering of charged
particles from an Aharonov--Bohm flux line for $\beta=\alpha^2$. The 
radial wave function is the Bessel function
\begin{equation}
A_m(k\rho)=\e^{i\dem}J_{\nu(m)}(k\rho) \;,
\end{equation}
where
\begin{equation}
\nu(m)=\sqrt{m^2+2\alpha m+\beta}\;, \label{eq:indx}
\end{equation}
so that the phase shifts have the form 
\begin{equation}
\dem=\big[|m|-\nu(m)\big]\pi/2 \;.
\label{eq:phaseshift}\end{equation}

In place of Eq.\ (\ref{eq:forcesquare}) we have
\begin{eqnarray}
\lefteqn{ \polarangle\bigg|\mepv{x}\bigg|^2 } \label{eq:vdpm}\\
& = & \frac{(2\pi)^3\pi}{m_0^2\vol^2}\sum_m\bigg|\int d\rho A^*_{m+1}
\bigg[\frac{\alpha}{\rho}\frac{\partial}{\partial\rho}+
\frac{\alpha m+ \beta}{\rho^2}\bigg]A_m\bigg|^2 \;. \nonumber
\end{eqnarray}
The integral on the right side of this equation can be evaluated in
terms of standard integrals of Bessel functions \cite{Gradshteyn80}, and is equal to
\begin{eqnarray}
\lefteqn{(k/\pi)\cos\big\{[\nu(m+1) -\nu(m)]\pi/2\big\}\e^{i(\dem-\dmp)}}
\label{eq:id1} \\
& \times &\frac{2\alpha[\nu(m+1)^2-\nu(m)^2-1]-4(\alpha
m+\beta)} {[\nu(m+1)^2-\nu(m)^2]^2-2\nu(m+1)^2-2\nu(m)^2 +1}\;. \nonumber
\end{eqnarray}
Substitution of Eq.\ (\ref{eq:indx}) into the fraction shows that it is
equal to unity, while Eq.\ (\ref{eq:phaseshift}) shows that the cosine
is equal to $\sin(\dmp-\dem)$ so that the force is given
in exactly the same form as in Eqs. (\ref{eq:2dlongi}) and (\ref{eq:tcs2d}).

For this type of potential the forward scattering amplitude diverges
like $1/\phi$, but for the transport cross section the factor of
$1-\cos\phi$ cancels the divergence.  For the transverse cross section
the situation is much more delicate, as is shown in the recent
discussion by Wexler and Thouless \cite{Wexler98}.

\section{Summary}

We have shown that, both for a short-ranged central potential, and for
a long-ranged potential of the sort given by a flux line or vortex, 
calculation of the longitudinal force on a moving potential in a
noninteracting background by kinetic theory and by  time-dependent 
perturbation theory give the same results to lowest order in the velocity.

We thank Ping Ao for helpful correspondence.
This work was partially support by NSF Grant No. DMR-9528345.

\appendix
\section{Radial Schr\"odinger Equation}

Here is the key identity for establishing the equivalence between the two theories.
Since we did not specify a particular pinning potential, any general property
of the matrix element could be derived from the radial wave equation
which in three dimensions is
\begin{equation}
\frac{1}{r}\frac{d^2}{dr^2}\big[rA_l(kr)\big]+\bigg[k^2-2m_0V(r)-\frac{l(l+1)}{r^2}\bigg]A_l(kr)=0 \;.
\end{equation}
A general relation between different $l$'s can be derived as following,
\begin{eqnarray}
\lefteqn{ r^n\bigg[A_\lp\frac{dA_l}{dr}-\frac{dA_\lp}{dr}A_l\bigg]_0^\infty = (\kp^2-k^2)\int drr^nA_\lp A_l }\nonumber\\
& + & (n-2)\int drr^{n-1}\bigg[A_\lp\frac{dA_l}{dr}-\frac{dA_\lp}{dr}A_l\bigg] \nonumber\\
& - & \Big[\lp(\lp+1)-l(l+1)\Big]\int drr^{n-2}A_\lp A_l \;. \label{eq:ra3}
\end{eqnarray}
To evaluate the radial part of the matrix element in Eq. (\ref{eq:pot}),
we first integrate it by parts, replace $V(r)A_l(kr)$ using the radial wave equation,
and then use Eq. (\ref{eq:ra3}) with $n=1$,
\begin{eqnarray}
\lefteqn{2m_0\int drr^2A_\lp\frac{dV}{dr}A_l = \bigg[r^2\frac{dA_\lp}{dr}\frac{dA_l}{dr}+r^2k^2A_\lp A_l\bigg]_0^\infty} \nonumber\\
&& \mbox{}+(\kp^2-k^2)\int dr\bigg[r^2A_\lp\frac{dA_l}{dr}-\frac{2l(l+1)}{\lp(\lp+1)-l(l+1)}rA_\lp A_l\bigg] \nonumber\\
&& -\bigg[\lp(\lp+1)-l(l+1)-2\frac{\lp(\lp+1)+l(l+1)}{\lp(\lp+1)-l(l+1)}\bigg]\int drA_\lp\frac{dA_l}{dr} \;. \nonumber\\
\end{eqnarray}
Using the asymptotic expansion for the Bessel functions, the boundary term gives the
difference between two phase shifts.
Similar relation could be established from the wave equation in two dimensions,
or could be done by directly substituting $A_l(x)\rightarrow\sqrt{\dfrac{\pi}{2 x}}A_m(x)$ with $l=m-1/2$,
\begin{eqnarray}
\lefteqn{ 2m_0\int d\rho\rho A_\mp\frac{dV}{d\rho}A_m = \bigg[\rho\frac{dA_\mp}{d\rho}\frac{dA_m}{d\rho}+\rho k^2A_\mp A_m\bigg]_0^\infty }\nonumber\\
& + & (\kp^2-k^2)\int d\rho\bigg[\rho A_\mp\frac{dA_m}{d\rho}-\frac{\mp^2+3m^2-1}{2(\mp^2-m^2)}A_\mp A_m\bigg] \nonumber\\
& - & \bigg[\mp^2-m^2-\frac{2\mp^2+2m^2-1}{\mp^2-m^2}\bigg] \nonumber\\
&& \makebox[1in]{}\hfill\int d\rho\bigg[\frac{A_\mp}{\rho}\frac{dA_m}{d\rho}-\frac{A_\mp A_m}{2\rho^2}\bigg] \;.
\end{eqnarray}

\end{multicols}

\end{document}